\documentclass[journal=jctcce]{achemso} 

\usepackage[utf8]{inputenc}
\usepackage{amsmath, amssymb, amsfonts}
\usepackage{titlesec}
\usepackage{hyperref}
\usepackage{mathrsfs}

\newenvironment{perspectivequote}[1]
  {%
   \def\quoteauthor{#1}
   \begin{quote}\centering\itshape\small%
  }
  {%
   \par\noindent\hfill\normalfont--- \quoteauthor
   \end{quote}%
  }

\newcommand{\Hil}{\mathscr{H}}

\title{After 100 Years of Quantum Mechanics:\\
Toward a Constructive Observation-Centered Perspective}

\author{Timothy Stroschein}
\affiliation{ETH Zurich, Department of Chemistry and Applied Biosciences, Vladimir-Prelog-Weg~2, 8093 Zurich, Switzerland}

\author{Markus Reiher}
\email{mreiher@ethz.ch}
\affiliation{ETH Zurich, Department of Chemistry and Applied Biosciences, Vladimir-Prelog-Weg~2, 8093 Zurich, Switzerland}

\begin{document}
\maketitle

\begin{abstract}
Quantum mechanics owes much of its extraordinary success to a
Hilbertian program of mathematical formalization. 
Yet, the
formalism remains poorly aligned with the practical limitations
of computations in finite dimensions and under finite accuracy. 
In this perspective, we argue that this mismatch points to the need for a 
new mathematical program: a rigorous
constructive theory for effective descriptions to identify
essential degrees of freedom. We propose an
observation-centered point of view in which signals are treated as the primary objects of analysis,
while wave functions and Hamiltonians are reconstructed as auxiliary structures to rationalize the observed data. 
Our starting point is
a signal-based spectral equation that reformulates frequency analysis
as an operator problem. 
We connect this point of view to results on prolate Fourier theory,
spectral analysis with finite observation time, and short-time quantum simulation. We 
highlight a sharp accuracy transition relating necessary observation time to the effective spectral density of
a signal for achieving accurate resolution. 
The resulting framework integrates approximation 
as a fundamental necessity
more directly into the foundations of quantum mechanics and points toward a broader program for the effective description of complex quantum systems, such as those found in the molecular sciences.
\end{abstract}

\section{Introduction}
In its description of nature, quantum mechanics has achieved an unprecedented 
formal clarity \cite{von_neumann_1927,von_neumann_1927_probability,von_neumann_1927_thermodynamics,neumann1929eigenwerttheorie}.
Its axioms are sharp, its mathematical language is precise, and its predictive 
power is unmatched.
Its success has fueled a century of optimism about the further reach
of mathematical principles in science, or, as Eugene Wigner had put it:

\begin{perspectivequote}{Wigner, 1960 \cite{Wigner59}}
The miracle of the appropriateness of the language of
mathematics for the formulation of the laws of physics is a wonderful gift which we neither
understand nor deserve. 
\end{perspectivequote}

Yet, even after a century, one central problem that remains
is that the equations 
cannot be solved in their exact form for realistic systems. Instead,
the computational treatment
of quantum mechanics relies on finite truncation, heuristic reduction and numerical compromise. 
The divergence between the exact formalism and practical computation is particularly severe in chemistry 
and materials science. While the underlying physical laws are widely regarded as known \cite{reiher_wolf_2009}, 
already Dirac pointed to the need for reliable, but feasible solution methods:

\begin{perspectivequote}{Dirac, 1929 \cite{dirac1929manyelectron}}
It therefore becomes desirable that approximate practical methods of applying quantum mechanics should be developed, 
which can lead to an explanation of the main features of complex atomic systems without too much computation.
\end{perspectivequote}

Dirac's call has been answered in practice with remarkable success, but the dimensionality 
of the resulting equations quickly becomes intractable.
Over the past century, fields of applied quantum mechanics such as quantum chemistry 
have developed an arsenal of approximate methods that has made quantum mechanics a predictive tool --- in chemistry, for molecular structure prediction, energy calculations, and spectroscopy \cite{cramer2004,jensen2017}.

However, what remains striking is that Dirac's call to develop efficient descriptions of larger systems has not yet matured into a comprehensive mathematical program pursued with the same rigor that shaped the formulation of quantum mechanics.
Instead, what exists today is a rich and successful collection of approximation methods, but not a unified theory of effective descriptions.
Hence, the difficulty today is not the absence of useful approximations, but the absence of a comparably coherent mathematical account of their validity and limitations.

What is often missing are rigorous error bounds.
For many effective descriptions, we still lack a quantitative understanding of how computational dimension, neglected degrees of freedom, and achieved accuracy are related.
Such bounds are not a technical luxury: they determine if a reduced model can be trusted, when 
agreement with experiment is meaningful, and how computational resources should be allocated to meet a prescribed accuracy target.

We believe that bringing to the application of quantum mechanics the same mathematical clarity that has shaped its formulation requires a different mathematical language: a constructive approximation theory.
The program we have in mind treats finite precision as fundamental and asks how reliable finite-dimensional descriptions can be derived with controlled accuracy and computational cost.

Much of the approximation theory we are referring to has already appeared in various forms and has long been applied.
Yet, one central question remains only rarely answered in a systematic way: how are dimensionality and achievable accuracy optimally related?
One field in which this question has received an unusually clear and deep answer is prolate Fourier theory.

In the early 1960s, H. Landau, H. Pollak, and D. Slepian asked how well
a band-limited function can be concentrated in a finite time interval.
This initiated a remarkably fruitful line of work and led to an important approximation theorem:
the space of band-limited functions that are also
concentrated in a time interval is, to high accuracy,
approximated by a space of dimension $2WT/\pi$ \cite{ProIII, OnBandwidth}.
Their series of seminal papers laid the foundation of
prolate Fourier theory \cite{ProI, ProII,ProIII, ProIV, ProV, OnBandwidth}.

Despite its potential and far-reaching implications for computation,
prolate Fourier theory has remained largely unnoticed in the natural sciences.
We do not believe that this neglect is accidental.
We argue that the slow development of a rigorous and much-needed approximation theory is tied to the architecture of quantum mechanics itself, which centers the description on infinite-dimensional and unobservable wave functions.
In this perspective article, we therefore seek to prepare the ground for a different view of quantum mechanics, one that is observation-centered and built around finite-precision data, much akin to an observer perspective in quantum information theory.

\section{The wave function formalism of quantum mechanics}

The time-dependent Schr\"odinger equation\cite{schrodinger1926eigenwertproblem_iv}
\begin{equation}
    i  \partial_t \Psi(t) = H \Psi(t)
    \label{eq:schrodinger}
\end{equation}
describes the time evolution of a quantum system by a
state vector $\Psi(t)$ in Hilbert space $\Hil$ generated by a
self-adjoint Hamiltonian $H$.
Ultimately, understanding the dynamics and physical properties of a system
amounts to solving the spectral decomposition of the Hamiltonian,
\begin{equation}
    H \varphi_n = E_n \varphi_n
    \label{eq:eigenvalue}
\end{equation}
Later developments, such as relativistic quantum mechanics and quantum electrodynamics, fit naturally 
into the Hilbert space formulation of quantum mechanics of state vectors subject to Schr\"odinger time evolution.
However, the computational application of the theory, which is key to reliable and meaningful applications in atomistic modeling and in the molecular sciences, forms a second
layer of complexity whose conceptual development is formally less clear, despite the successes of numerical techniques developed in quantum chemistry in the past 80 years. One must choose a
representation, select an expansion basis, truncate it, evaluate matrix
elements approximately, control numerical conditioning, and decide
whether the resulting approximation is accurate enough for the problem
at hand. These steps do not share the formal exactitude of the theory
itself. Yet, they are essential for confronting an abstract theory with
experimental data and thereby attributing physical relevance to it and turning it into a reliable modeling approach.

If the precise formulation and its approximate application are viewed as
parts of a single theory, the focus changes. It is no longer central
whether a theory is strictly exact, but whether it yields effective
descriptions with controlled computational cost and controlled accuracy.
Approximations are unavoidable in both, experimental
measurement and computational prediction. The real task is therefore
to reconcile the illuminating clarity of formalized mathematics with practical
approximation.

What remains striking is that this second layer has never matured into a
comparably coherent formal theory --- despite the remarkable successes of numerical approaches developed in quantum chemistry.
Instead, it has advanced largely through
isolated approximation schemes rather than a unified mathematical
architecture. We attribute part of this stagnation to a structural
weakness of quantum mechanics, already emphasized by von Neumann early on:

\begin{perspectivequote}{von Neumann, 1927 \cite{von_neumann_1927}}
    A common deficiency in all of these methods is, however, that they introduce elements into the calculation that are
    in principle unobservable and physically pointless.\\
    \dots it is, however, unsatisfactory and unclear why this detour through the unobservable and the non-invariant is necessary.\footnote{Authors' translation from the German original: ``Ein gemeinsamer Mangel aller dieser Methoden ist aber, dass sie prinzipiell unbeobachtbare und physikalisch sinnlose Elemente in die Rechnung einf\"uhren. \dots es ist aber unbefriedigend und unklar, weshalb der Umweg durch das nicht-beobachtbare und nicht invariante notwendig ist.''}
\end{perspectivequote}

There is little hope of rigorously deriving effective finite-dimensional descriptions
of the salient features of a system if one begins by axiomatizing
wave functions 
as infinite-dimensional and unobservable objects 
central to the description.
Moreover, the wave function formalism carries a substantial unitary redundancy:
unitarily equivalent representations encode the same physics.
In this sense, the theory is burdened with degrees of freedom that 
carry no physical information, yet may entail computational overhead.

Of course, avoiding the wave function for the reasons given above has been a driving force behind the development of density functional theory (DFT), where the electron density alone (or including additional, similarly instructive fields such as the spin density \cite{JacobReiher2012})
should be sufficient to obtain an electronic energy \cite{HohenbergKohn1964}.
However, despite the proven exactness of DFT \cite{Lieb1983}
and the tremendous success of approximate Kohn-Sham DFT as a computational tool in chemistry, materials science, solid state physics, and molecular biology, it has remained a rather ad-hoc approach with substantial semi-empirical character in practice, unsuitable for the rigorous framework that we envision here for the whole of quantum mechanics. 
As a side remark, we note that it has not even been possible to equip approximate DFT with reliable system-specific error assessment,
and therefore, much of it has been Bayesian in nature \cite{Mortensen2005,SimmReiher2016}
or based on benchmarking \cite{Mardirossian2017,Goerigk2022}.

All of this motivates the perspective adopted here. Rather than taking
wave functions as the primary objects of analysis, we shift the focus to
signals. Signals encode the spectral information from
which a quantum description can be reconstructed. 
This does not mean abandoning the wave function formalism, but rather
treating it as secondary to the quantities actually accessed in
experiment, much in the spirit of high-resolution
molecular spectroscopy, where effective Hamiltonians are reconstructed
from measured transitions \cite{field2015spectra,lefebvrebrion_field2004,lefebvrebrion_field1986,Quack1990,QuackMerkt2011}.
Starting from signals leads to a
more constructive theory, which integrates finite precision more closely into the foundation and highlights a more comprehensive role for
approximation theory.

\section{The observation-centered perspective}

Depending on the protocol, experiments probe autocorrelation functions,
expectation values, transition amplitudes, dipole correlations, survival
probabilities, response functions, and related signals.
Many such signals admit a representation of the form
\begin{equation}
    S(t) = \sum_k \alpha_k e^{i \omega_k t},
    \label{eq:signal}
\end{equation}
with amplitudes $\alpha_k \in \mathbb{C}$ and frequencies $\omega_k \in \mathbb{R}$. 
Continuous spectral contributions can be accommodated by replacing the sum with an integral.
The frequencies $\omega_k$ carry information about energies or, more
typically, energy differences, while the amplitudes $\alpha_k$ encode
overlaps, transition strengths, or matrix elements. Their precise
interpretation depends on the experiment, but the general mathematical
structure remains the same.

For example, if $\Psi(t)$ solves the Schr\"odinger equation and $O$ is an observable, then
\begin{align}
    S(t) & = \langle \Psi(t), O \Psi(t) \rangle\\
         & = \sum_{k,l}
            \underbrace{\langle \Psi(0), \varphi_k \rangle
            \langle \varphi_k, O \varphi_l \rangle
            \langle \varphi_l, \Psi(0) \rangle}_{\alpha_{kl}}
            e^{i(E_k - E_l)t}.
    \label{eq:observable_signal}
\end{align}
Likewise, the autocorrelation function
\begin{equation}
    S(t) = \langle \Psi(t), \Psi(0) \rangle
         = \sum_k \underbrace{\left| \langle \varphi_k, \Psi(0) \rangle \right|^2}_{\alpha_{k}} e^{iE_k t}
\end{equation}
directly reveals the energy components present in the state.

This is the starting point of the observation-centered perspective that
we propose. Rather than taking the wave function formalism as fundamental
and treating signals as derived quantities, we begin with a family of
signals $\{S_i(t)\}$ associated with a physical system and ask how much
of the underlying quantum description can be reconstructed from them in
a controlled and computationally efficient way.

This reverses the usual logic. Wave functions and Hamiltonians are no
longer introduced as an axiomatic foundation, but as auxiliary structures  
that account for observed data and support controlled extrapolation
beyond them. The central task is therefore to extract spectral
information directly from the data themselves. The goal is not merely to
reproduce observations, but to obtain theories whose predictions remain
reliable beyond the measured data.

To extract the spectral information encoded in signals in a way that
can subsequently support the construction of effective quantum
descriptions, we place the following spectral equation 
proposed in Ref. \citenum{stroschein2025groundexcitedstateenergiesanalytic}
at the center of
the analysis :
\begin{equation}
   - i \partial_t (S \ast f)(t) = \omega (S \ast f)(t),
   \label{eq:signal_spectral_equation}
\end{equation}
where $f \in L^2(\mathbb{R})$ is a test function, $S \ast f$ denotes
the convolution of $f$ with the signal $S$, 
and $\omega \in \mathbb{R}$ is an eigenvalue to be
identified.

If the signal admits the decomposition of Eq.~\eqref{eq:signal}, then
substituting it into Eq.~\eqref{eq:signal_spectral_equation} yields
\begin{equation}
    \sum_k \omega_k \alpha_k F(\omega_k)e^{i\omega_k t}
    =
    \omega \sum_k \alpha_k F(\omega_k)e^{i\omega_k t},
\end{equation}
where $F$ denotes the Fourier transform of $f$. 
In particular, the only admissible eigenvalues are frequencies
$\omega_n$ appearing in $S$. For a nontrivial solution, the Fourier transform of the test function
must have $F(\omega_n)\neq 0$ and
$F(\omega_k)=0$ for all $k\neq n$.

Therefore, Eq.\,\eqref{eq:signal_spectral_equation} recasts frequency
analysis as an operator-based spectral problem derived directly from the
observed signal. Its major advantage is that finite measurement constraints can be
incorporated directly into the spectral problem, allowing us to derive
fundamental approximation bounds.

\section{Approximating the spectral equation with error bounds}

One fundamental constraint, investigated extensively in
Refs.\,\citenum{stroschein2024prolatespheroidalwavefunctions,stroschein2025groundexcitedstateenergiesanalytic},
is the finite observation time of the signals.
This constraint can be studied rigorously by restricting the test functions to the space of time-limited functions.
Within this space, prolate spheroidal wave functions provide an optimal expansion basis for reducing the operator equation to a finite-dimensional matrix problem.

Building on the core insight of prolate Fourier analysis, we obtain a sharp resolution threshold\cite{stroschein2025groundexcitedstateenergiesanalytic}:
accurate reconstruction becomes possible once the available observation time is large enough relative to the effective spectral density, namely when
\begin{align}
    \delta_{\mathrm{eff}} \lesssim \frac{T}{\pi}.
    \label{eq:transitions}
\end{align}
Here $\delta_{\mathrm{eff}}$ denotes the approximate number of frequencies per unit bandwidth in the spectral region of interest.
Once the observation time exceeds this threshold, the errors in the recovered frequencies decay exponentially fast.
In particular, $T$ does not need to be asymptotically large, but can remain of similar order as $\delta_{\mathrm{eff}}$.
By contrast, standard Fourier analysis requires observation times asymptotically larger than the reciprocal of the smallest relevant frequency separation.

The sharp, non-asymptotic transition in accuracy encoded in Eq. \eqref{eq:transitions} singles out observation time as a fundamental computational resource:
the denser the relevant spectrum, the longer one must observe the signal to resolve it accurately.
This makes Eq. \eqref{eq:transitions} a useful building block for composite approximation schemes and resource allocation, including quantum computation as discussed in Section~\ref{sec:QPD} below.

Ref.\,\citenum{stroschein2025groundexcitedstateenergiesanalytic}
considers the practical restriction that only discrete signal samples
$S(t_k)$ are available. In this setting, the prolate sampling formula for band-limited signals \cite{walter_sampling_2003}
provides a natural discretization of the overlap integrals from which
the matrix elements are computed. Its main advantage over the classical
Whittaker--Shannon interpolation formula \cite{NoiseCommunication} lies in its substantially
better truncation behavior.
Ref.\,\citenum{stroschein2024prolatespheroidalwavefunctions} combines
this formula with new concentration identities for prolate spheroidal
wave functions to obtain improved truncation estimates and, in
particular, a $2WT/\pi$ sampling theorem: $W$-band-limited signals that
are strongly concentrated in the interval $[-T,T]$ can be reconstructed
to high accuracy from $2WT/\pi$ sampling points.

Remarkably, the  discretization error exhibits the same sharp
accuracy transition \eqref{eq:transitions} that already arose from the
finite-time restriction and off-band frequencies. This shows that the two approximation subroutines are aligned: the combined scheme balances different
sources of error such that they are of similar order of magnitude. 

Although the exposition in these references has focused on signals arising as quantum
autocorrelation functions, the resulting approximation scheme extends
readily to arbitrary signals of the form given in Eq.~\eqref{eq:signal}.

These examples do not exhaust the finite-accuracy constraints that can be
incorporated into Eq.\,\eqref{eq:signal_spectral_equation}. Beyond the finite-time and sampling
restrictions analyzed in the references cited above, one may also account for
errors in data acquisition by introducing an additive perturbation to the
signal,
$$ \tilde S(t) = S(t) + n(t). $$
The structure of the noise $n(t)$ reflects its origin: if
$\tilde S(t)$ arises from simulated time evolution, the numerical error
may accumulate with time, so that the magnitude of $n(t)$ grows with
$t$, whereas in the presence of statistical noise the sampled values
$n(t_k)$ can be modeled as random variables. In
either case, Eq.\,\eqref{eq:signal_spectral_equation} remains a principled
starting point for deriving 
approximation schemes with rigorous error
bounds 
on frequency and amplitude estimates.

To treat such finite-accuracy constraints within a unified approximation
theory, a general framework for subspace methods in
spectral analysis can be developed (see Ref.
\citenum{stroschein2025approximationframeworksubspacebasedmethods}). 
The focus shifts from a specific algorithm to the general approximation
problem underlying many methods: approximating local spectra of
self-adjoint, possibly unbounded operators through a finite expansion
space.
The framework introduces an error measure that originates 
from the projection-valued measure and remains well-defined even for self-adjoint 
unbounded operators, where standard matrix-norm or residual-based estimates cease to apply.

Building on this measure, we
derive in Ref. \citenum{stroschein2025approximationframeworksubspacebasedmethods}
integrated spectral inequalities that separate the intrinsic
subspace approximation error from additional perturbations arising from
noise, discretization, or other numerical procedures.
This enables a unified and non-asymptotic quantification of multiple error sources. 
The new framework also 
provides a rigorous clarification of frequent 
numerical artifacts such as spectral pollution and spurious eigenvalues.
Dimension detection 
in the presence of noise, provides an algorithmic solution to such artifacts and a 
starting point to derive stability guarantees. 

More broadly, this framework supports the central ambition of the approximation program advocated here: 
approximation should not enter quantum mechanics as a mere collection of ad hoc numerical tricks. 
Instead, it calls for a coherent theory of how spectral information is compressed, how errors propagate, and 
how computational dimension governs achievable accuracy.

\section{An application: quantum computation}
\label{sec:QPD}

While the approximation framework presented here applies to both experiment and classical calculations, quantum computation provides a particularly natural setting to advance the approximation program.
In quantum simulation, the unitary time evolution of a quantum
system is implemented on a quantum computer, while spectral information
is extracted from the resulting dynamics. In 
quantum phase estimation, both the time evolution and the spectral
estimation are performed on the quantum device. More recently,
however, a class of hybrid algorithms has emerged in which only the time
evolution is carried out on the quantum device, while the spectral
analysis is delegated to a classical post-processing routine
\cite{klymko2022uvqpe,ding_lin_2023_qcels,ding_et_al_2024_qmegs,shen_et_al_2025_odmd}.

Within this class of hybrid algorithms, simulation time on the quantum
device becomes the primary measure of computational cost. The central
question is therefore what determines the simulation time required to
obtain accurate energy estimates. In
Ref.\,\citenum{stroschein2025groundexcitedstateenergiesanalytic}, we therefore
developed quantum prolate diagonalization (QPD) as a hybrid algorithm
designed to attain an optimal relationship between finite simulation
time and achieved accuracy. Our rigorous error bounds and the accuracy
transition \eqref{eq:transitions} show that the spectral density
in the region of interest is the key quantity governing the required
simulation time and, more broadly, the optimal allocation of quantum
resources.

This application to hybrid quantum algorithms and
finite simulation time is only one manifestation of a more general
framework \cite{stroschein2025approximationframeworksubspacebasedmethods}.
Additional constraints, such as imperfect time evolution and statistical
noise in signal samples, can be incorporated seamlessly into the same
analysis. 

\section{Conclusions}

The extraordinary success of quantum mechanics over the past century was
driven in no small part by a Hilbertian program of mathematical
formalization \cite{hilbert1902mathematical_problems,vonNeumann1927Beweistheorie,von_neumann_1932,birkhoff_von_neumann_1936}. Its axiomatization in terms of Hilbert spaces,
self-adjoint operators, and spectral theory gave physics a language of
remarkable precision and intellectual eloquence. This achievement shaped the appearance of modern 
mathematics to describe physics and hence also theoretical chemistry. It remains one of the greatest successes in the history of science.

Still, the formalization program reveals its limitations when confronted with the
more complex layers of the natural world. Formal exactness alone does
not tell us which degrees of freedom are essential, how finite-precision
predictions should emerge from exact equations, or how complexity should
be reduced without sacrificing predictive relevance. In this regime, the
central challenge is no longer to formulate exact theories, but
to understand how effective descriptions arise while controlling accuracy
and computational cost \cite{Reiher2022persp}.

Therefore, we have called in this perspective for a new program to develop a rigorous approximation 
formalism that appears so necessary in understanding the 
more messy and complex layers of nature:
a program that seeks to understand how 
consistent theories emerge from observations of finite precision.
As a starting point, we have
introduced a signal-based spectral equation to extract 
spectral information directly from observed data. 
In this way, 'approximation' enters not
as a collection of ad-hoc numerical tricks, but as a part of the
foundation from which a consistent theory is constructed.

This program is not purely aspirational. It has already led to 
results of far-reaching implication.
By applying prolate Fourier theory 
to the signal-based spectral equation, we have derived 
an accuracy transition that precisely relates 
spectral density to necessary observation time \cite{stroschein2024prolatespheroidalwavefunctions,stroschein2025groundexcitedstateenergiesanalytic}.
The non-asymptotic accuracy transition can now be used to optimize the allocation 
of computational resources in composite approximation schemes.
Such relationships between physical information, 
computational resources and achievable accuracy are 
exactly the kind of structural laws we hope to uncover.
While we are currently working on applications in different areas of quantum chemistry, we have already demonstrated the value of the new framework in devising a minimal-resources quantum algorithm for energy measurements on a quantum computer \cite{stroschein2025groundexcitedstateenergiesanalytic}.

Much remains to be developed, but if the last 100 years were shaped by the
formalization of quantum theory, the next may require an equally serious
mathematical effort toward effective descriptions. 
We hope that the present perspective helps articulate this program 
and invites others to participate in its expansion.

\section*{Acknowledgments}
The authors gratefully acknowledge financial support from the Swiss National Science Foundation (project no. 200021\_219616) and from the Novo Nordisk Foundation (Grant No. NNF20OC0059939 'Quantum for Life').

\providecommand{\refin}[1]{\\ \textbf{Referenced in:} #1}

\end{document}